\documentclass[twocolumn,prd,aps,tightenlines,floats,floatfix,preprintnumbers,nofootinbib,eqsecnum]{revtex4-2}

\def\sp{\kern +3pt}
\def\sm{\kern -3pt}
\def\spQ{\kern +6pt}

\def\bea{\begin{eqnarray}}
\def\eea{\end{eqnarray}}

\def\sfrac#1#2{{\textstyle \frac{#1}{#2}}}

\def\be{\begin{equation}}
\def\ee{\end{equation}}
\def\ba{\begin{eqnarray}}
\def\ea{\end{eqnarray}}

\usepackage{graphics}
\usepackage{graphicx}
\usepackage{epsf}
\usepackage{amsmath}
\usepackage{amssymb}
\usepackage{xcolor}

\usepackage{ulem}

\begin{document}

\phantom{0}    
\vspace{-0.2in}  
\hspace{5.5in}

% include preprint number option
\preprint{{\bf LFTC-26-07/116}}

\vspace{-1in}%\parbox{1.5in}{ \vspace{-9.6in}}  % moves the preprint box down

\title
    {\bf Local-nuclear-density dependent calculation of nucleon electromagnetic form factor ratios in finite nuclei} 
\author{G.~Ramalho$^1$, 
K.~Tsushima$^2$ and Myung-Ki Cheoun$^1$}
\vspace{-0.1in}

\affiliation{$^1$Department of Physics and OMEG Institute, 
Soongsil University, \\
Seoul 06978, Republic of Korea
\vspace{-0.15in}}
\affiliation{$^2$Laborat\'orio de 
  F\'{i}sica Te\'orica e Computacional -- LFTC,
  Programa de P\'{o}sgradua\c{c}\~{a}o em Astrof\'{i}sica e F\'{i}sica Computacional,
  Universidade Cidade de  S\~ao Paulo, 
01506-000,   S\~ao Paulo, SP, Brazil}

\vspace{0.2in}
\date{\today}
\phantom{0}

\begin{abstract}
We calculate the electromagnetic form factors of nucleons bound in finite nuclei
and make predictions to the ratio between the electric ($G_E^\ast$)
and magnetic ($G_M^\ast$) form factors in terms of the square
of the four-momentum transfer $Q^2$.
We extend our previous constant-density calculations within the
covariant spectator–QMC framework by incorporating the spatial nuclear density profiles
of finite nuclei and quantify the deviations from
the average-density approximation used in our previous applications.
The impact of the medium effects can then be observed 
considering the ratio between $G_E^\ast/G_M^\ast$ and the ratio in free space $G_E/G_M$,
that define the proton electromagnetic double ratio associated with a given nucleus.
The double ratio is expected to remove some systematic uncertainties.
There is the expectation that ratios $G_E^\ast/G_M^\ast$ associated with the bound protons inside
the nucleus will be measured in the near future in polarization-transfer experiments
$(\vec{e} A, e'\! A' \vec{p})$.
Anticipating future measurements on the subject, we make predictions for the double ratios,
to be compared with the average measurements on the energy states of protons bound to nuclei.
We consider the nuclei $^{12}$C, $^{16}$O and $^{40}$Ca.
We conclude that the form factors calculated using nuclear density profile function
$\rho(r)$ are less suppressed than in the case of the results calculated using the average nuclear density. 
The results indicate that the relative importance of the low-density
surface region increases with $Q^2$, leading to a weaker suppression
than that predicted by the average-density approximation.
We also make predictions for the ratios associated with neutrons bound to nuclei.
\end{abstract}

%\phantom{0}
%\vspace{7.0in}
%\vspace{-6in}
\vspace*{0.9in}  % sets how far the title is below the preprint box
\maketitle

\section{Introduction \label{secIntro}}

The structure of hadrons and the properties of the degrees of freedom of QCD, are expected to
be modified in nuclear matter~\cite{Brown91a,Symmetry2025,Saito07a,GA-medium,Medium1}.
Evidence of the medium modifications of electromagnetic and axial form factors of the nucleon in
terms of the squared transfer momentum $q^2$ (or $Q^2=-q^2$) has been observed in experiments
with nuclei~\cite{Brown85,Gysbers19,Strauch03a}.

The experimental information about the in-medium electroweak form factor of baryons is very scarce.
There is then a strong motivation to develop theoretical models that can be used to estimate
structure functions of baryons at relatively high densities and moderate and large $Q^2$, based on
the degrees of freedom manifest in vacuum: valence quarks and meson cloud excitations of baryon
cores~\cite{Symmetry2025,Saito07a,GA-medium,Medium1}.
Model calculations of the electromagnetic form factors (EMFFs) of the baryons
in a nuclear medium predict sizable deviations from the calculations in the free
space~\cite{Frank96a,Lu99a,Lu98a,Smith04a,Cloet09a,Araujo18a}.

A very promising technique to measure the medium modifications of the EMFFs of the baryons is
the polarization-transfer method~\cite{Strauch03a,Dieterich01,Paolone10a,JLabPAC35}.
The method has been used to determine the ratio between the electric and magnetic form factors of
the proton in electron-proton scattering with polarized electrons, by measuring the ratio between
the transverse and longitudinal polarizations of the proton in the final state~\cite{Puckett17a}.
The extension of the method for the neutron has also been proposed for studies at Jefferson
Lab~\cite{JLabPAC35}.
For the proton, the proposal is to consider quasi-elastic electron-nucleus scattering where a proton
is removed from the nucleus~\cite{Strauch03a,Dieterich01,Paolone10a}.
The ratio between the transverse and longitudinal polarizations of the proton in the final state
can be used to determine the ratio  $G_E^\ast/G_M^\ast$ of the bound proton (we use $\ast$ to
represent variables and functions in-medium)~\cite{Strauch03a}.

The measurements of the ratio $G_E/G_M$ in free space and in medium $G_E^\ast/G_M^\ast$ provide
then a direct quantification of the impact of the medium on the EMFFs of the
proton~\cite{Medium1,Strauch03a,Medium2}.
These measurements have been, however, restricted almost exclusively to the protons bound to the
nucleus $^4$He~\cite{Strauch03a,Dieterich01,Paolone10a}.
Partial measurements have been performed using $^{12}$C and $^{40}$Ca at
MAMI~\cite{Izraeli18a,Bricelji20a,Kolar20a,Kolar24a,Kolar23a,Paul20a}.
Experiments with $^{16}$O were performed at JLab~\cite{Malov00a}.
Contrarily to the experiments with $^{4}$He [$^{4}$He breakup: $^{4}$He$(\vec{p}, e' \vec{p})^3$H]
where the kinematics can be reduced to an almost elastic scattering, and $Q^2$ is the single
variable to be considered, the electron-nucleus scattering with a nucleus composed of several
protons is more complex.
In the case of the nucleus there are additional degrees of freedom.

In the electron-nucleus reactions the measured polarizations are not exclusive functions of $Q^2$.
In that case, one needs to take into account additional dependencies associated with the
quasi-elastic reactions, namely on the three-momentum of the bound proton, and the bound energy of
the probed proton~\cite{Strauch03a,Dieterich01,Izraeli18a}.
For each energy state of the bound proton the polarization variables depend on $Q^2$, and on the
missing momentum of the proton, that can be expressed in terms of
the {\it off-shellness} variable $\nu$ (or virtuality)~\cite{Paolone10a,Izraeli18a,Kolar23a}.
In the present work, we focus on the global modification of the EMFFs and perform a spatial average over the total nuclear density distribution, without resolving the missing-momentum dependence or the individual proton energy states.

In previous works, we had studied the EMFFs of nucleons and hyperons in a nuclear
medium~\cite{Symmetry2025,Medium1,GA-medium,Medium2}, using a covariant quark model based on valence
quarks and the meson cloud excitations of the baryon cores, combined with the formalism of the
quark-meson coupling (QMC) model~\cite{Saito07a,Lu99a,Lu01a,Lu98a,Saito96a}.
In those studies we considered the baryons immersed in an infinite nuclear matter, characterized by a constant density $\rho$.
The method has been used to calculate the electroweak form factors of the octet baryons, from low
densities (close to the vacuum, $\rho=0$), to
  densities associated with finite nuclei ($\rho=0.5$--0.8$\rho_0$), up to high densities
$\rho= 3\rho_0$~\cite{Symmetry2025}, where $\rho_0 = 0.15$ fm$^{-3}$ is the normal nuclear matter density.
The latter case can be used in applications for astrophysical implications.

In the previous applications to finite nuclei~\cite{Symmetry2025}, we considered
the average density associated with the nucleus, based on the nuclear density
profile function $\rho(r)$ as calculated by the QMC model~\cite{Saito96a}.
The function $\rho(r)$ represents the combination of proton and neutron densities,
also known as the local baryon density, not the nucleon charge density~\cite{Lu99a,Lu98b}.
In symmetric nuclear matter the proton and neutron densities are equivalent.
It has been argued that the use of the average density is not a realistic approximation, since
it ignores the surface effects and should therefore be avoided in calculations related to
interactions with baryons.

Local-density treatments of the EMFFs of bound nucleons in finite nuclei
have been considered in earlier QMC studies. In particular, Lu {\it et al.}~\cite{Lu99a}
calculated the form factors of protons occupying specified shell-model orbits
by folding the density-dependent nuclear-matter form factors with
the corresponding single-particle proton densities.
A related prescription was applied to the neutron charge form factor in $^{3}$He~\cite{Lu98b}.
Density-dependent form factors have also been incorporated
into reaction calculations of polarization-transfer observables,
including shell-dependent effective densities
and final-state interactions~\cite{Kolar24a,Ron13a}.
These calculations differ from the average-density treatment adopted
in our recent applications and provide an important precedent
for the use of spatially varying densities.
The purpose of the present work is therefore not to introduce the local-density concept itself,
but to implement it using the density-dependent EMFFs 
of the covariant spectator–QMC framework and to quantify the accuracy
of the average-density approximation for nucleus-averaged proton and neutron EMFF ratios.

In the present work, we extend our covariant spectator–QMC formalism from nuclear matter
with a constant density to finite nuclei described by
a position-dependent density $\rho (r)$, and compare the resulting
EMFFs with those obtained using the average density.
The central objective is to quantify the accuracy of the average-density approximation
and to determine how its validity changes with $Q^2$ and with the nucleus under consideration.

Aiming for the comparison with experimental data, we make predictions for
the nucleon EMFFs associated with some middle mass nuclei and for the averaged double ratios
$(G_E^\ast/G_M^\ast)/(G_E/G_M)$.
We consider in particular the nuclei $^{12}$C, $^{16}$O and  $^{40}$Ca, since they have been used
conveniently on electron-nucleus scattering experiments.
This comparison allows us to separate two related questions: the overall medium modification
generated by the average nuclear density and the additional correction associated
with the spatial variation of the nuclear density. The latter correction,
rather than the average-density prescription itself, is the main subject of the present work.
Our calculations cannot at the moment be directly compared with the measurements of the double
ratios, because the measurements are presently mainly focused on the study on the {\it off-shellness}
variable~\cite{Izraeli18a}.

Our predictions are obtained considering the averages on the energies and positions of the bound
nucleons probed by the scattering with the electron beam.
For the calculations we use the formalism developed previously for the baryon octet in Refs.~\cite{Medium1,Medium2}, restricted in the present work to protons and neutrons.
In a first step, we restrict our analysis of the EMFFs
$G_E^\ast$ and $G_M^\ast$ to the case of the bound protons.
At the end, we also present estimates for the case of the
bound neutrons in order to infer the expected changes from the
proton to the neutron cases, anticipating the possibility of experimental measurements
for bound neutrons~\cite{JLabPAC35}.

We expect that our predictions may be tested in the near future in experiments at MAMI (low-$Q^2$)
and JLab (intermediate- and large-$Q^2$), when accumulated data on the {\it off-shellness} variable
became available.
From our calculations, we conclude that our estimates are consistent with the available
measurements for $^{12}$C and $^{40}$Ca below $Q^2 = 0.5$ GeV$^2$.
We conclude also that the mean-field approximation (average density) is a good approximation
for the electric and magnetic form factors of the  bound proton below $Q^2 = 0.25$ GeV$^2$.
The effect of the nucleus surface on the electric and magnetic form factors (deviation from model
with average density) may start to be pronounced only for values of $Q^2$ above 1 GeV$^2$.
The impact of the shape of the nuclear density on the double ratio $(G_E^\ast/G_M^\ast)/(G_E/G_M)$
is, however, small.
Measurements in the intermediate-$Q^2$ region, particularly above $Q^2 \simeq$ 2 GeV$^{2}$,
would be useful for testing the deviations between the density-profile and average-density calculations.

The present article is organized as follows:
In the next section we discuss the polarization-transfer method used in the study of the proton
elastic form factors and electron-nucleus scattering to probe the electromagnetic structure of
the bound protons.
In Sec.~\ref{sec-mod-medium} we review briefly the formalism used for the calculation of
the EMFFs of the nucleons in free space, and in a nuclear medium with constant baryon density.
The extension of the formalism to finite nuclei is discussed in Sec.~\ref{secFF-medium}.
The numerical calculations for the EMFFs of bound nucleons are presented and
discussed in Sec.~\ref{sec-nucleus}.
The outlook and conclusions are given in Sec.~\ref{sec-conclusions}.

\section{Polarization-transfer method for bound nucleons
  \label{secPolarization}}

The polarization-transfer method was originally developed to measure the electromagnetic ratio
$G_E/G_M$ of the proton~\cite{Puckett17a}. 
There are ongoing studies on the extension of the method
for the neutron using deuterium targets~\cite{JLabPAC35,Madey03a}.
It has been suggested that the methodology can also be used to probe the electromagnetic structure
of protons using quasi-elastic reactions with a nucleus.
The idea is that in reactions where a proton is knocked out from a nucleus the
out coming proton carries the information of the bound proton~\cite{Medium1,Medium2,Strauch03a}.

For the experiments with free protons we can express the ratio between transverse polarizations,
labeled as $P_x$, and the longitudinal polarizations, labeled as $P_z$,
as~\cite{Puckett17a,Izraeli18a}
\ba
R \equiv \frac{P_x}{P_z} = - \frac{2M}{(E+ E') \tan \frac{\theta_e}{2}}
\frac{G_E}{G_M},
\label{eqRatio}
\ea
where $M$ is the proton mass, $E$ is the incident electron energy, $E'$ is the energy of the
scattered electron, and $\theta_e$ is the scattering angle between the electrons (assuming elastic
electron-proton scattering).

The relation (\ref{eqRatio}) can then be used to measure the ratio $G_E/G_M$ for protons in free
space.
When we consider the electron-nucleus reactions, one needs to consider quasi-elastic
electron-proton scattering.
In this case the function $R$ has additional dependencies on the energy of the bound proton
and the {\it off-shellness}~\cite{Paolone10a,Izraeli18a,Kolar23a}.
In comparison to the elastic electron-proton scattering one needs then to consider two additional
degrees of freedom: the continuous dependence on the {\it off-shellness} (missing momentum) and on
the proton energy state
(the orbital energy in a shell model: $s$, $p$, $d$, ...).

For the study of the medium modification of the bound proton EMFFs one can
consider the double ratio
between the ratio $R$ for the scattering with a nucleus $A$: $R_A$,
and the ratio to the proton case (hydrogen nucleus H):
$R_H$ to determine the ratio between the electromagnetic ratios $G_E/G_M$ between free protons and
bound protons  $G_E^\ast/G_M^\ast$  (associated with the polarizations $P_x^\prime$ and $P_z^\prime$
in the medium)~\cite{Strauch03a,Paolone10a,Izraeli18a,Ron13a}:
\ba
\frac{R_A}{R_H} = \frac{(P_x^\prime/P_z^\prime)}{(P_x/P_z)}
\simeq \frac{G_E^\ast/G_M^\ast}{G_E/G_M}.
\label{eqRatioA}
\ea
The second equality is valid only if the kinematics of the elastic scattering (function $R_H$) and
the quasi-elastic scattering (function $R_A$) are similar. The approximate relation is applicable when the elastic and quasi-elastic kinematic factors are sufficiently similar and when the residual reaction effects are small or largely canceled in the ratio.

In the following, we refer to the r.h.s.~of Eq.~(\ref{eqRatioA}) as the
EMFF double ratio. It should be distinguished from a complete reaction calculation
of the polarization-transfer double ratio on the l.h.s.
The approximate correspondence assumes that the relevant kinematic factors
and residual nuclear effects largely cancel between the nuclear and proton targets.
The present work calculates the EMFF ratio and does not explicitly evaluate
final-state interactions, two-body currents, missing-momentum dependence,
or experimental acceptance.

The advantage of the polarization-transfer experiments is that the polarization ratios can be
determined with a very good precision, since the systematic uncertainties are canceled on the
double ratios, allowing accurate determinations of the in-medium EMFF
ratio~\cite{Puckett17a,Izraeli18a}.
In reactions with a nucleus (quasi-elastic scattering) it is also necessary
to take into account many-body effects related with the structure of the nucleus,
including meson-exchange currents, isobar configurations and final-state interactions
(FSI)~\cite{Malov00a,Dieterich01,Ron13a}.
The relative contributions to these effects are, in general, small~\cite{Strauch03a,Paolone10a}.
The contributions of the FSI may be more significant, but the effects are reduced when we
consider the double ratios $R_A/R_H$~\cite{Paolone10a,Strauch04a}.
The conclusion is then that the comparison between models and data may be hard when we look for the
individual polarizations ($P_x$, $P_z$, $P_x^\prime$ and $P_z^\prime$), but it is simplified when we
consider ratios between polarizations~\cite{Strauch03a,Izraeli18a,Ron13a}.

The relation (\ref{eqRatioA}) has been used to determine the double ratio
$(G_E^\ast/G_M^\ast)/(G_E/G_M)$
for the  $^{4}$He~\cite{Strauch03a,Paolone10a,Medium1}.
In the case of the $^{4}$He, only one energy level for the bound proton (ground state) is
considered in the analysis.
For the cases under discussion in the present article ($^{12}$C, $^{16}$O and $^{40}$Ca), we also
assume that the modifications on the multiplicative factor of Eq.~(\ref{eqRatio}) are small, and we
use the ratio between polarizations in Eq.~(\ref{eqRatioA}) to determine the double ratio
$(G_E^\ast/G_M^\ast)/(G_E/G_M)$.
Studies on the deuterium ($^2$H), where the proton is weakly bound are consistent with the $G_E/G_M$
results of protons in the free space~\cite{Paul19a,Izraeli18a,Ron13a}.

Contrary to other works that take into account many-body effects,
in the present work we do attempt to describe the function $R_A/R_H$ in terms of the virtuality
$\nu$, and the energy of the bound proton.
We aim at the description of the global effect given by the average of the function
$\frac{R_A}{R_H}$ on $\nu$, and the combined effect of the different proton energy states
(considering the average on the states).
Using this procedure (average on virtuality and proton energy states)
we make predictions to the average double ratio $(G_E^\ast/G_M^\ast)/(G_E/G_M)$ in terms of $Q^2$.
For that purpose, we take into account the nucleon density of the nucleus $A$, using the nuclear
distribution $\rho(r)$ determined by the QMC model~\cite{Saito07a,Saito96a}.
The nuclear density profile function is associated with the average of the possible energy
states of the bound proton.
In fact, all the wave functions belong to all the nucleon shells in a nucleus
summed up to get the nuclear density at the point $r$ of the nucleus.

\section{Formalism for the nuclear medium \label{sec-mod-medium}} 

For the calculations of the EMFFs of the nucleon in free space and in
nuclear matter, we combine the frameworks of the covariant spectator quark model and the QMC model
as described in Refs.~\cite{Symmetry2025,GA-medium,Medium1,Octet}.

The covariant spectator quark model is a constituent quark model where the baryons are described as
three-constituent quark systems~\cite{Symmetry2025,NSTAR2017,Octet}.
In the framework the baryon is composed of a spectator quark-pair and a single quark that is free
to interact with electroweak fields in the impulse approximation, while the interactions with the
quarks are described in terms of quark electroweak form factors that simulate the structure
associated with the gluon and quark-antiquark dressing of the quarks~\cite{Nucleon,Omega}.
The constituent quark form factors are parametrized using a vector meson dominance picture
calibrated in the study of the electroweak structure of the nucleon, the baryon octet, and the
baryon decuplet~\cite{Octet,Nucleon,Omega,Axial}.
In the calculations of the transition current between baryon states, one can reduce the baryon wave
function to a quark-diquark wave function, where the radial part can be determined
phenomenologically by experimental data or lattice QCD results for some ground state
systems~\cite{Octet,Omega}.

Since the contributions of the valence quarks are, in general, not sufficient to describe the
electromagnetic and axial structure of the baryons and transitions between baryon states at low
$Q^2$, the model considers in some cases also the contributions associated with the baryon-meson
states in a phenomenological form~\cite{Octet,Medium1}.

The covariant spectator quark model~\cite{Nucleon,Omega,NSTAR2017} was originally developed for the
study of the nucleon EMFFs and transitions between the nucleon and the nucleon resonances in
the spacelike region~\cite{PPNP2024,NSTAR,Nstar1,Nstar2,Roper}.
The framework has been extended to the SU(3) sector and applied to the study
of the EMFFs of octet baryons, decuplet baryons, and other baryon
systems~\cite{Omega,OctetDecuplet}.
The formalism has also been used in the study of baryon transitions in the timelike
region~\cite{NstarTL}, in the electroweak structure of baryons in free space~\cite{Axial},
and the nucleon deep inelastic scattering~\cite{nucleonDIS}.

We discuss now the formalism associated with the octet baryon, in general,
and the nucleon, in particular.
For the calculation of transition form factors one has then to consider the parameters associated
with the valence quark structure, including the radial wave functions, and some coefficients
associated with the vector meson dominance parametrization of the quark form factors.
All the parameters were determined in previous works~\cite{Octet,Medium1}.
For the description of the meson cloud contributions, one needs to consider in addition 5
coefficients and 2 cutoffs for the description of the octet baryon EMFFs in free
space~\cite{Medium1}.

In the following, we use $G_\ell^\ast$ ($\ell =E,M$) to represent the electric ($\ell =E$) and
magnetic ($\ell =M$) form factors.
The octet baryon EMFFs are calculated in the form ($\ell =E,M$)
\ba
G_\ell^\ast (Q^2) = Z_B \left[ G_\ell^{{\rm B} \ast} (Q^2) +
  G_\ell^{{\rm MC} \ast} (Q^2)
  \right], 
\ea
where $G_\ell^{{\rm B} \ast}$ represents the contribution from the bare core (valence quarks) and
$G_\ell^{{\rm MC} \ast}$ the contribution associated with the meson cloud (MC).

The contributions for the valence quarks are determined by the extension of the model for the
nucleon~\cite{Nucleon} to the baryon octet~\cite{Octet}.
The meson cloud contributions for the form factors $G_\ell^{{\rm MC} \ast}$ are estimated based on
the baryon-meson dressing diagrams, where there is a coupling with the intermediate baryons, and
the coupling with the mesons~\cite{OctetMM}.

In the case of the nucleons, the meson cloud contributions are dominated by $\pi N$
states~\cite{OctetMM}.
For the proton, one needs to consider the $\pi^+ n$ and $\pi^0 p$ contributions.
As for the neutron, one needs to consider the $\pi^0 n$ and $\pi^- p$ contributions.
The conclusion is then that the calculation of the meson cloud contributions to the proton form
factors requires the calculation of the proton and neutron bare contributions.
Similarly, the proton and neutron bare contributions are also necessary for the calculation of the
neutron meson cloud contributions.

For the extension of the formalism to a nuclear medium with constant baryon density, for the bare
contribution, we consider the model parametrization of the covariant spectator quark model
for the octet baryons, with the baryon and meson masses replaced by the effective
masses determined by the QMC model~\cite{Saito07a,Saito96a}.
As for the meson cloud contribution, we take into account the medium modifications on the
meson-baryon coupling constants, as determined by the QMC model result of axial couplings and
hadron masses.
The formalism and expressions used in the calculation of electroweak form factors in a nuclear
medium with constant baryon density can be found in Refs.~\cite{Symmetry2025,Medium1,GA-medium}.
The local-density folding procedure used in the next section is conceptually related to earlier QMC calculations of orbit-dependent bound-nucleon form factors. The elementary density-dependent form factors employed here are nevertheless different: they are obtained from the covariant spectator quark model, including the valence quark core and meson cloud contributions, combined with the in-medium masses and couplings determined by the QMC model. The present calculation therefore provides a finite-nucleus application of this specific form factor framework over an extended range of $Q^2$.

\begin{figure*}[t]
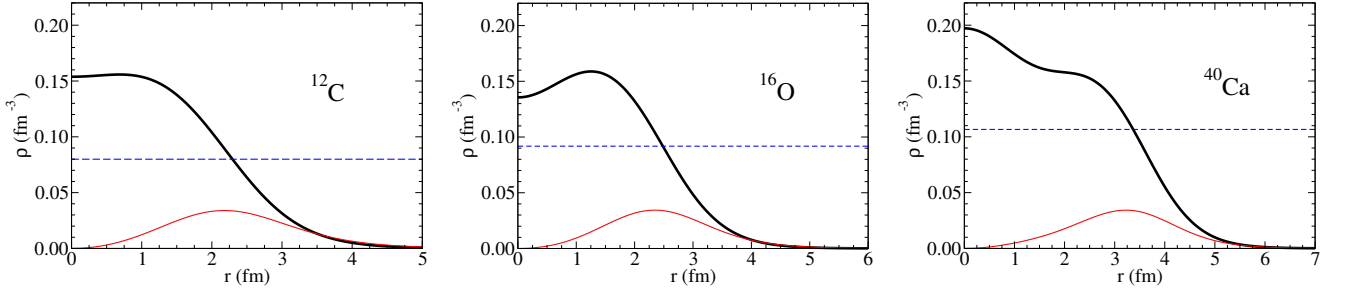
   %%%%    Figure 1 (page 5)    
\vspace{.3cm}
\begin{center}
\mbox{
  \includegraphics[width=2.2in]{rho-12C-mod1} \hspace{.1cm}
  \includegraphics[width=2.2in]{rho-16O-mod1} \hspace{.1cm}
\includegraphics[width=2.2in]{rho-40Ca-mod1} }
\end{center}
  \caption{\footnotesize
  Nuclear density profile functions for the nuclei $^{12}$C, $^{16}$O and $^{40}$Ca.
  The thick solid line represent $\rho(r)$ from the QMC model from Ref.~\cite{Saito96a}.
  The dashed line represents the average density $\bar \rho_A$.
  The thin solid line represents the function $g(r)$ in relative units, discussed in the main text.
\label{fig-rho-r}}
\end{figure*}

\section{Electromagnetic form factors of nucleons  bound to a nucleus -- formalism
\label{secFF-medium} }

We discuss how the EMFFs obtained in nuclear matter with
a constant density are applied to nucleons bound in a finite nucleus.
The starting point is the model developed for symmetric nuclear matter
with constant density, as described in Refs.~\cite{Medium1,Medium2}. 
The local-density prescription itself has precedents in earlier finite-nucleus QMC studies;
here it is used to assess the difference between the full density-profile calculation
and the average-density approximation within the present covariant spectator–QMC framework.

Next, we explain how the model is extended for the calculation of the EMFFs
for the case where the nuclear density changes with the position according to a profile density
function $\rho(r)$, where $r= |{\bf r}|$ is the distance to the center of the nucleus $A$.
The function $\rho (r)$ depends only on the radial variable.
Given a nuclear density function $\rho(r)$, we can calculate the average density $\bar \rho_A$
for the nucleus $A$, using $\rho$ as a weight function:
\ba
\bar \rho_A = {\cal Z}^{-1} \int_0^{\infty}
\left[ \rho(r) \right] \rho(r) \, r^2 dr,
\ea
where
\ba
   {\cal Z} =  \int_0^{\infty} \rho(r) \, r^2 dr.
   \label{eqZnorm}
\ea
The normalization factor ${\cal Z}$ introduced in Eq.~(\ref{eqZnorm})
should not be confused with the proton number.
It denotes only the radial normalization integral.
The function $\rho(r)$ is normalized according to
\ba
4 \pi {\cal Z} = A,
\ea
where $A$ is the mass number of the nucleus $A$,
the sum of the number of protons ($Z$) and neutrons ($N$).
The factor $4 \pi$ takes into account the angular integration.
Recall that $\rho(r)$ contains the nucleon density coming from all
the nucleon shell wave functions of protons and neutrons.

The description of the EMFFs of nucleons immersed in finite nuclei requires the knowledge of
the form factors $G_\ell^\ast$ in terms of $Q^2$, and the density $\rho(r)$, that varies with the
variable $r$, or $\rho(r)$.
For the calculation of the effective in-nucleus form factors $\overline{G_\ell^\ast}$ we consider
the local-density approximation, where the contribution associated with a nucleon at the distance
$r$ from the center of the nucleus is determined by the nuclear density function
$\rho(r)$~\cite{Lu99a,Lu98b,Bremsstrahlung} 
\ba
\overline{G_\ell^\ast} (Q^2) & =&
     {\cal Z}^{-1} \int_0^{\infty} \left[r^2 \rho(r) \right] G_\ell^\ast
     \left(Q^2, \rho(r) \right)  dr \nonumber \\
     & = & \frac{4 \pi}{A} \int_0^{\infty} \left[r^2 \rho(r) \right] G_\ell^\ast
     \left(Q^2, \rho(r) \right) dr,
     \label{eqGl-medium}
\ea
where ${\cal Z}$ is also defined by Eq.~(\ref{eqZnorm}).
For the $N=Z$ nuclei considered here,
we can well approximate identical normalized proton and neutron spatial distributions,
$\rho_p (r) / Z =\rho_n(r)/ N = \rho_B (r) /A$. The total baryon density
can therefore be used as the averaging weight for both the proton and neutron form factors.

In the present work, the weight function $\rho(r)$ represents the average nucleon density
associated with the relevant mean fields for the bound proton states, since we are
focused on the global medium modifications associated with the nuclei.
The study of the contributions of the different energy states is possible within a microscopic
model that calculates the wave functions associated with the energy levels of the bound
nucleons~\cite{Lu99a,Malov00a}.

The numerical calculation of $\overline{G_\ell^\ast} (Q^2)$ can be a complicated task for fast
varying functions, but is simplified for finite nucleus, when the nuclear density functions
are moderately varying functions, and the values become negligible after
a given value of $r = \bar R$.
The numerical calculations can be further simplified when we decompose the nuclear distribution
functions into intervals where the functions are almost linear.

The nuclear density profile functions for $^{12}$C, $^{16}$O and $^{40}$Ca are presented in
Fig.~\ref{fig-rho-r}.
In general, the nuclear density profile function can be divided into three sub-regions:
the central part with high densities ($\rho = 0.8$--$1.2 \rho_0$), the intermediate region, where
we can observe the smooth falloff of the function, and the surface region with a slow transition to
the free space density ($\rho \simeq 0$).
In the figures, we also include the density associated with the average density (horizontal dashed
line) for each nucleus.

When we consider a formalism with constant densities,
one can estimate the nucleon form factors associated with the nucleus
by $G_\ell^\ast \left(Q^2,\bar  \rho_A \right)$, where $\bar  \rho_A$ is the average density of the
nucleus $A$.
This result can also be denominated as the mean-field approximation,
where the density is assumed to be constant on a large volume.

The calculation based on Eq.~(\ref{eqGl-medium}) is more realistic than the mean-field
approximation, since it takes into account the variation of the nuclear density function with $r$.
In that case we take into account not only the high density region of $\rho(r)$ near the center,
but also the intermediate and surface regions, associated with the transition to the vacuum, where
the form factors can be approximated by the free space form factors.

For the application to the nuclear medium it is important to quantify the difference
between the calculations that take into account the shape of the nuclear density function:
$\overline{G_\ell^\ast} (Q^2)$, and the calculations based on the mean-field approximation:
$G_\ell^\ast \left(Q^2,\bar  \rho_A \right)$.
The comparison between the two methods can be made within our framework if we use
the same functions  $G_\ell^\ast \left(Q^2,\rho_i \right)$ to calculate $G_\ell^\ast \left(Q^2,\bar
 \rho_A \right)$ and to calculate the integral (\ref{eqGl-medium}).

\subsection{Numerical calculation of the functions \label{secGell}}

From the observation of Eq.~(\ref{eqGl-medium})  one can conclude that the contribution of the
constant density $\rho_i$ is determined by the function $ g(r) = r^2 \rho (r)$ for the point
associated with the density $\rho_i$.
The function $g(r)$ is also represented in Fig.~\ref{fig-rho-r} by the thin solid line.

The radial weight $g(r)=r^2 \rho (r)$ generally reaches its maximum in a region where the local density is comparable to the average density 
${\bar \rho}_A$. Thus, the intermediate-density region dominates the normalization-weighted average, whereas the low-density tail becomes increasingly important for the $Q^2$ dependence of the result. This observation explains why the average-density approximation can provide a reasonable first estimate, particularly at low $Q^2$. Contributions from lower and higher densities, however, remain present but can become relevant when their associated form factors have different $Q^2$ dependencies.

For the numeric calculation of the function $\overline{G_\ell^\ast} (Q^2)$, through the integral
(\ref{eqGl-medium}), we consider a finite grid of points $R_i$ associated with the density $\rho_i =
\rho(R_i)$.
We use a grid of $n + 1$ points ($n$ intervals), where the last point ($R_{n+1}$) has a density
$\rho_{n+1} = \rho(R_{n+1})$ that is negligible when compared with the densities near the center
$\rho(R_1) \approx \rho_0$.
In these conditions the region $r > R_{n+1}$ can be approximated by $\rho(r) \simeq 0$, and ignored
in the calculation of the numerical integrals.
In practice, the last point of the grid depends on the nucleus under discussion, and should be
chosen case by case.

Given a grid defined by the intervals $[R_i , R_{i+1} ]$ ($i =1, .., n$), we can decompose the
integral (\ref{eqGl-medium}) into a sum of $n$ integrals.
Using the midpoint approximation for each interval, one can write
\ba
\overline{G_\ell^\ast} (Q^2) \simeq \sum_{i=1}^n a_i G_\ell^\ast \left(Q^2,\bar \rho_i \right),
\label{eqGl-medium2}
\ea
where $\bar \rho_i$ is the average of the endpoints of the interval $\bar \rho_i =
\sfrac{1}{2}(\rho_i + \rho_{i+1})$, and
\ba
&& a_i = \frac{{\cal Z}_i}{\cal Z}, \label{eqai} \\
&& {\cal Z}_i = \int_{R_i}^{R_{i+1}} \rho (r) \, r^2 dr,  \label{eqZi}\\
&& {\cal Z} = \sum_{i=1}^n  {\cal Z}_i. 
\ea
From the normalization of the function $\rho(r)$ one can conclude that $\sum_i a_i =1$.
Details about the numerical calculation of the integrals and the midpoint approximation can be
found in Appendix~\ref{appendix-p1}.

The relation (\ref{eqGl-medium2}) provides accurate calculations for the EMFFs if the
functions are well approximated by linear functions in the intervals $[R_i , R_{i+1}]$.
In these conditions, we can estimate the in-medium form factors considering a sum of a few terms
associated with a few different values for the nuclear density $\bar \rho_i$.
In addition, the magnitude of the coefficients $a_i$, that can be determined analytically, provides
relevant information about the weight associated with the region of densities $\bar \rho_i$.

In the next section we present the numerical calculations of the electric and magnetic form factors
of bound nucleons.

\section{Electromagnetic form factors of the nucleons bound to a nucleus  \label{sec-nucleus}}

In this section we present and discuss our numerical results for
the EMFFs of the bound nucleon in nuclei $^{12}$C, $^{16}$O and $^{40}$Ca.
We consider the nuclear density distributions represented in
Fig.~\ref{fig-rho-r}. Our primary interest is not only the magnitude of the medium modification relative to the free-space form factors, but also the difference between the two finite-nucleus prescriptions. The comparison between the solid and dashed curves therefore provides a direct measure of the correction to the average-density approximation generated by the spatial nuclear density profile.

Our calculations for the ratios  $G_E/G_M$ and $G_E^\ast/G_M^\ast$, and 
the double ratios $(G_E^\ast/G_M^\ast)/(G_E/G_M)$ are presented in
Figs.~\ref{fig-12C}, \ref{fig-16O} and \ref{fig-40Ca} for
$^{12}$C, $^{16}$O and $^{40}$Ca, respectively.
The results for $G_E^\ast$ and $G_M^\ast$ are normalized by the functions
in free space ($G_E$ and $G_M$ respectively) for an easy
visualization of the in-medium effects. The double ratios
$(G_E^\ast/G_M^\ast)/(G_E/G_M)$ would be the unit (constant one) if there are no
medium modifications.

In the figures, we present the calculations that take
into account the shape of the nuclear density profile function $\rho(r)$,
given by the function  $\overline{G_\ell^\ast} (Q^2)$ (solid lines), and the result
of the mean-field approximation given by the calculations
with the (constant) average density $\bar \rho_A$, as determined by
the function $G_\ell^\ast \left(Q^2,\bar \rho_A \right)$ (dashed lines).

In the calculations we use $\bar \rho_A  = 0.533 \, \rho_0$ for the $^{12}$C, $\bar \rho_A  = 0.612
\, \rho_0$ for the $^{16}$O, and $\bar \rho_A = 0.711\, \rho_0$  for the $^{40}$Ca, based on the
nuclear densities functions from Ref.~\cite{Saito96a}, presented in Fig.~\ref{fig-rho-r}.

For the comparison of the properties of the nuclei $^{12}$C, $^{16}$O and $^{40}$Ca, we include in
Table~\ref{tab-radius} the expected values for the double ratios $(G_E^\ast/G_M^\ast)/(G_E/G_M)$ for
values of $Q^2$ from $0.5$ to 2.5 GeV$^2$ with steps of 0.5 GeV$^2$.

%  HVspaces

\vspace{-.5cm}

\subsection{Electric and magnetic form factors in a nucleus}

The calculations for  $G_E^\ast/G_E$ and $G_E^\ast/G_M$ are presented in the upper and middle
panels from Figs.~\ref{fig-12C}, \ref{fig-16O} and \ref{fig-40Ca}, for the nuclei $^{12}$C,
$^{16}$O and $^{40}$Ca, respectively.

The density-profile and average-density calculations give very similar results at low momentum transfer.
For all three nuclei, the solid and dashed curves are practically indistinguishable below
$Q^2 \simeq 0.2$--0.3 GeV$^2$. The average-density approximation is therefore accurate in this region.
At larger $Q^2$, the two calculations gradually separate, with the density-profile
form factors falling more slowly than the corresponding average-density results.

The relative differences in the calculations of $G_E^\ast$ and $G_M^\ast$ based on the nuclear
density function and in the mean-field approximation are presented in Table~\ref{table-diffs}, in
percentages, for several values of $Q^2$. To quantify the central result of the present work, Table~\ref{table-diffs} presents the relative deviations of the density-profile calculation from the average-density approximation.
Notice, how the differences are small for $Q^2 = 0.5$ GeV$^2$, and how the differences increase for
$Q^2 > 1.5$ GeV$^2$, for different nuclei.
The shape of the nuclear density profile function has a stronger impact on the electric form
factors than on the magnetic form factors. At $Q^2 =$ 2.5 GeV$^2$, the deviations in $G_E^\ast$ reach 4.3\%, 6.3\%, and 8.1\% for $^{12}$C, $^{16}$O, and $^{40}$Ca, respectively. The corresponding deviations in $G_M^\ast$ are smaller. This result demonstrates that the nuclear density profile has a larger impact on the electric form factor and that the correction generally becomes more significant for denser nuclei.

We discuss now why the form factors are less suppressed (weaker quenched effect) when we take into
account the shape of nuclear density $\rho(r)$, and the falloff in the intermediate region, for
large values of $Q^2$.
This effect can be understood from the analysis of Eq.~(\ref{eqGl-medium2}).

It is known that the contribution from the high densities (first terms) is associated with faster
falloffs, and that the densities associated with last terms (tail terms) have slower falloffs
(similar to those in free space).
The intermediate density contributions (densities in the range 0.5--0.7$\rho_0$) are
determined by densities close to the average density.
The main difference to the estimate based on the average density and the estimate
(\ref{eqGl-medium2}) is that in the last case there are contributions with faster and slower
falloffs, than the form factors with the average density.
The larger magnitude observed for the combined form factors is then the outcome of the dominance of
the low density contributions ($\rho \approx 0$) over the large density contributions ($\rho \approx
\rho_0$).
The reduction of the magnitude of the form factors determined by the nuclear density compared with
the mean-field approximation is then the consequence of stronger contributions from the outer region
(tail of nuclear density function) compared with the form factors estimated by the average density.
The exact calculation includes explicit contributions from the surface region which contribute to a
reduction of the form factors in absolute values.
The results indicate two distinct regimes. At low momentum transfer,
$Q^2 \lesssim 0.25$ GeV$^{2}$, the density-profile and average-density calculations are nearly indistinguishable. In this region, the average-density approximation is therefore sufficient within the numerical accuracy of the present calculation. At larger $Q^2$, the form factors associated with the higher-density regions decrease more rapidly, while the low-density components remain closer to their free-space behavior. Consequently, the relative importance of the nuclear surface increases, and the density-profile form factors become less suppressed than the average-density estimates.

\subsection{Electromagnetic double ratios in a nucleus}

We discuss now the double ratios $(G_E^\ast/G_M^\ast)/$ $(G_E/G_M)$.
Recall that the double ratios are, at the moment, the observables that can be directly accessed by
polarization-transfer measurements, as discussed in Sec.~\ref{secPolarization}.
For the direct comparison with the data, we multiply our results by $C_N= \frac{M_N^\ast}{M_N}$,
in order to convert the magnetic form factors in units of the nuclear magneton in a nuclear medium
($e/(2 M_N^\ast)$), into the units of the nuclear magneton in the free space ($e/(2 M_N)$).
A more detailed discussion of this effect can be found in Ref.~\cite{Symmetry2025},
and in Cloet {\it et al.}~\cite{Cloet09a}.

The quenched effect has observed in previous works for the proton double ratios in
a nuclear medium with constant density~\cite{Medium1}, where the quenching
effects are observed as an almost-linear reduction with $Q^2$,
when we consider the variation of the density inside the nucleus.
The global effect is a partial reduction of the quenching, a trend that
increases with $Q^2$.

The comparison of the results for $G_E^\ast$, $G_M^\ast$ and
the ratios $G_E^\ast/G_M^\ast$ based on the nuclear density profile function
and in the mean-field approximation, is presented in Table~\ref{table-diffs}, in percentage.
Notice how the relative differences for $G_E^\ast/G_M^\ast$
are reduced compared with the results for $G_E^\ast$.

\begin{figure}[t]
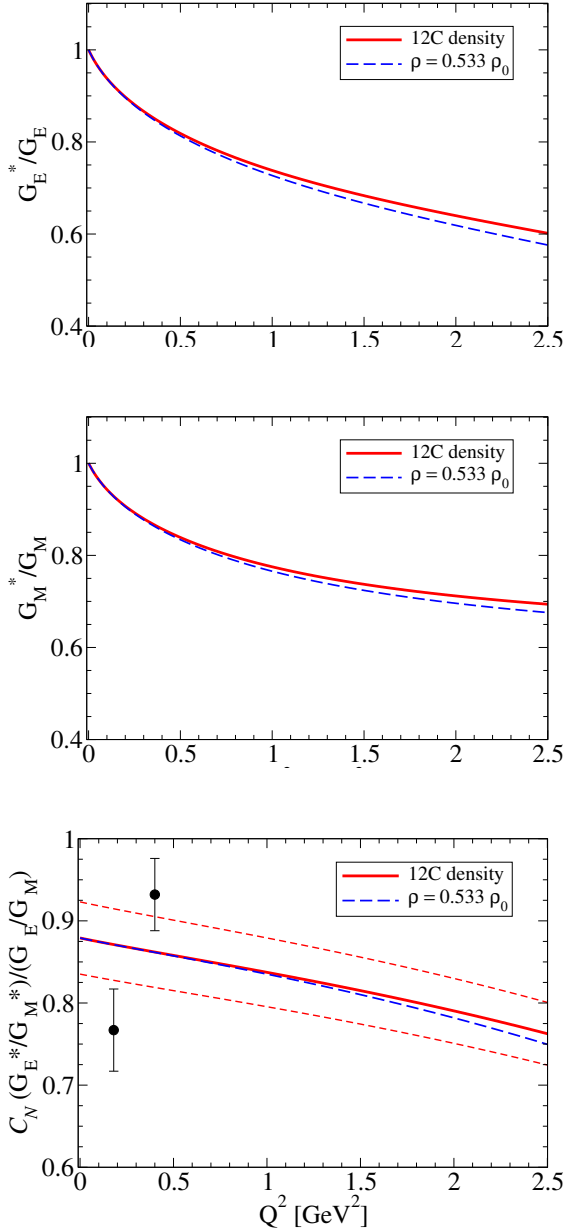
     %%  Figure 2  (page 7)
  \vspace{.3cm}
  \begin{center}
    \includegraphics[width=2.9in]{GE-12C}
  \end{center}
  \begin{center}
    \includegraphics[width=2.9in]{GM-12C}
    \end{center}
  \begin{center}
    \includegraphics[width=2.9in]{GEGM-12C}
  \end{center}
  \caption{\footnotesize EMFF of the bound proton in $^{12}$C. The data are
    from Ref.~\cite{Izraeli18a}.
    The solid line is the calculation based on the density-profile function and,  the thick-dashed line
    is the calculation based on the constant density.
    For the double ratios the thin-dashed curves
    indicate an variation of $\pm 5\%$ on the density-profile calculation.
\label{fig-12C} }
\end{figure}
\begin{figure}[t]
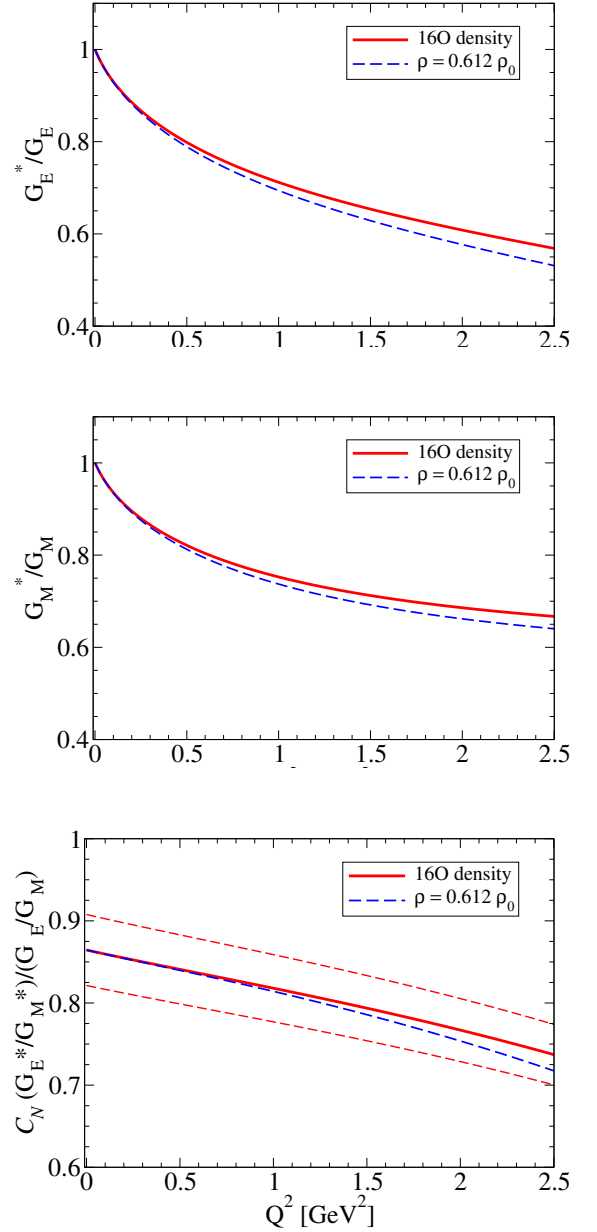
     %%  Figure 3  (page 7)
  \vspace{.3cm}
 \begin{center}
    \includegraphics[width=2.9in]{GE-16O}
  \end{center}
  \begin{center}
    \includegraphics[width=2.9in]{GM-16O}
    \end{center}
  \begin{center}
    \includegraphics[width=2.9in]{GEGM-16O}
  \end{center}
  \caption{\footnotesize EMFF of the proton bound to the nucleus $^{16}$O.
    Labels as in Fig.~\ref{fig-12C}.
\label{fig-16O} }
\end{figure}

It is worth noticing that the quenched effect observed
on the form factors $G_E^\ast$ and $G_M^\ast$ is partially canceled
when we calculate the double ratios.
This effect has also been observed in other works~\cite{Lu99a,Smith04a,Cloet09a,Araujo18a}.
t is a consequence of the correlated quenched effect on $G_E^\ast$ and $G_M^\ast$.
The simplest illustration of this effect comes from the observation that the electric square charge
radius and the magnetic dipole square radius for the proton are enhanced, and that
$G_E^\ast/G_M^\ast \propto 1 - \sfrac{1}{6}(r_E^{\ast \, 2} - r_M^{\ast \, 2}) Q^2$,
below $Q^2=1$ GeV$^2$, as discussed in Ref.~\cite{Cloet09a}.
There is a partial cancellation between the effects for
the magnetic and electric form factors (reduction of combined square radius
$r_E^{\ast \, 2} - r_M^{\ast \, 2}$), but the reduction is more significant in a nuclear medium
than in free space (enhancement of $r_E^{\ast \, 2} - r_M^{\ast \, 2}$), leading to the quenching
of $G_E^\ast/G_M^\ast$.

\begin{figure}[t]
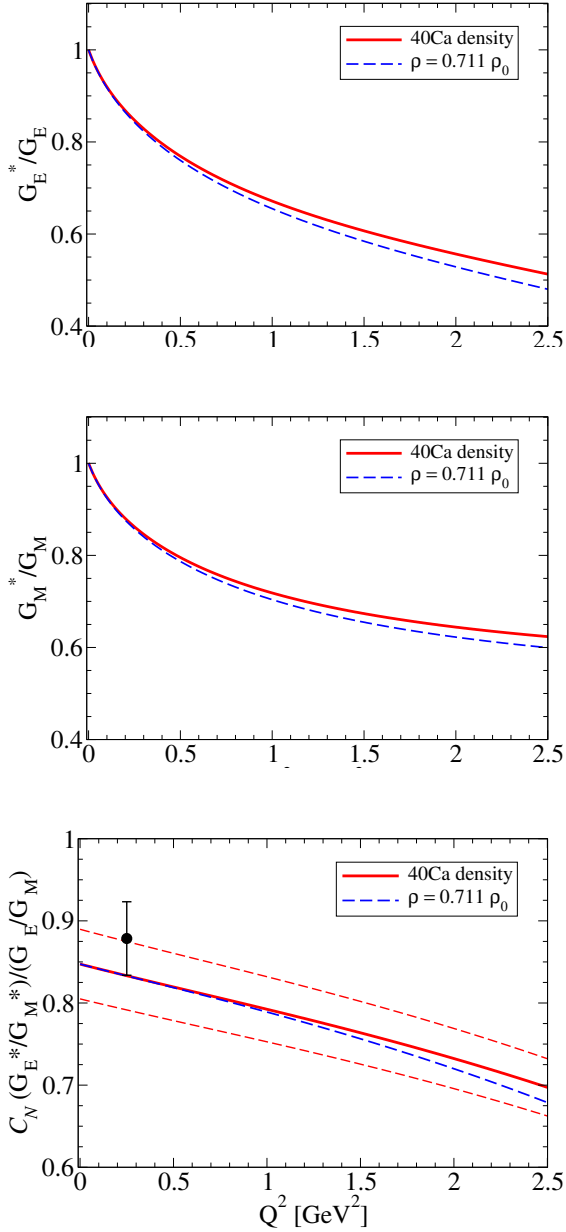
   %%  Figure 4 (page 8)
  \vspace{.3cm}
  \begin{center}
    \includegraphics[width=2.9in]{GE-40Ca}
  \end{center}
  \begin{center}
    \includegraphics[width=2.9in]{GM-40Ca}
    \end{center}
  \begin{center}
    \includegraphics[width=2.9in]{GEGM-40Ca}
  \end{center}
  \caption{\footnotesize  EMFF of the bound proton in $^{40}$Ca.
    The data are from Ref.~\cite{Kolar23a}.
    Labels as in Figs.~\ref{fig-12C} and \ref{fig-16O}.
\label{fig-40Ca}}
\end{figure}

There are also estimates of nucleon EMFF based on QMC model~\cite{Lu98a,Lu98b},
chiral-Soliton/Skryrme models~\cite{Frank96a,Christov95a,Yakhshiev03a}, and vector meson dominance
models~\cite{Rawat26a}, for different densities.

The quenching of the double ratios has been observed in the measurements of the proton double
ratios for $^{4}$He~\cite{Strauch03a}.

Although there are no data for the electromagnetic double ratios, except for the case of the
$^{4}$He, we make an effort to compare our model calculations with the data obtained for transverse
longitudinal polarization ratios for $^{12}$C and $^{40}$Ca at MAMI A1~\cite{Izraeli18a,Kolar23a}.
For the comparison with our calculation we use the relation (\ref{eqRatioA}), derived under the
assumption that the multiplicative factor between the polarization ratio $P_x /P_z$ and the ratio
between form factors $G_E/G_M$ [see Eq.~(\ref{eqRatio})] is almost unmodified from the free space
experiments to the experiments with nuclear targets.
For the explicit comparison with the data we consider a 5\% band of variation
(thin-dashed lines) associated with the theoretical uncertainties of relation (\ref{eqRatioA}).

Since we assume that our model calculations take into account all possible states of the bound
proton, and that the integration on the densities includes contributions of the protons in all
possible positions (from near the center to the outer region of the nucleus), for the
comparison with the polarization-transfer data, we combine (average) all possible virtuality values
and consider (averaging) the possible energy of the bound protons.
The data displayed in the figures are calculated from the
$(P_x^\prime/P_z^\prime)/(P_x /P_z)$ data for different values of missing proton momentum, combining
the results from different energy states ($s$, $p$, $d$, ...).

In the case of $^{12}$C there are data for two values of $Q^2$ ($Q^2 = 0.18, 0.40$
GeV$^2$)~\cite{Izraeli18a}.
The estimate for the double ratios is determined using 19 data points for the polarization ratios
in terms of $\nu$.
The final result includes the average on the states $1 s_{1/2}$ and $1 p_{3/2}$~\cite{Izraeli18a}.
As for the $^{40}$Ca, there are data only for $Q^2 = 0.25$ GeV$^2$~\cite{Kolar24a}, corresponding
to 13 $\nu$ data points associated to the proton energy states: $2 s_{1/2}$, $1 d_{3/2}$, and
continuous~\cite{Kolar24a}.

Notice in Fig.~\ref{fig-12C}, for $^{12}$C that the 2 data points are not too far away from our
estimated band of variation.
More measurements of the polarization-transfer double ratios for a wider range of $Q^2$ can further
test our calculations.
Future data can be used to test if the quenching effect is in fact stronger for larger values of
$Q^2$, and if the large suppression observed for $Q^2 = 0.18$  GeV$^2$ is a correct trend, or the exception.

Future measurements for larger values of $Q^2$ may help to test our model calculations for larger
values of $Q^2$, and in particular, those above $Q^2 = 2$  GeV$^2$,
can be used to confirm if the shape of the nuclear density profile function is relevant or
not (difference between solid and dashed lines). An important result is that the density-profile correction is considerably smaller in the double ratio than in the individual electric form factor. Since both $G_E^\ast$ and $G_M^\ast$ become less suppressed when the density profile is included, part of the correction cancels in $G_E^\ast/G_M^\ast$. Thus, the average-density approximation may be insufficient for precision calculations of the individual form factors at intermediate $Q^2$, while it remains more accurate for the proton double ratio.

\subsection{Remarks about the precision of the calculations}

To finish the discussion about EMFFs of the bound proton, we
discuss the numerical precision of the calculations based on the methods discussed in
Appendices~\ref{appendix-p1} and \ref{appendix-p2}.
The EMFFs are determined using the numerical integration based on finite grids of positions
and densities.

The number of points considered for each case was adjusted in order to have a precision better than
0.25\% in the calculation of the normalization of $\rho(r)$ based on the relation
$A = 4 \pi {\cal Z}$, and the same precision in the calculation of the average density
$\bar\rho_A$.
One considers then 7 points for the $^{12}$C, 12 points for the $^{16}$O, and 10 points for
$^{40}$Ca (see Appendix~\ref{appendix-p1}).
As before, we define $n$ intervals associated with $n+1$ points.

In a first stage, we calculate the function $\overline{G_\ell^\ast} (Q^2)$ using the midpoint
method considering $n$ densities.
To test the precision of the calculations, in a second stage, we recalculate the form factors
including the endpoints of the original grids, obtaining a calculation with more $n+1$ points.
To improve the accuracy of the numerical calculations of the integrals, in the second stage, we
consider the method of Simpson (see Appendix~\ref{appendix-p2}).

Combining the precision associated with the normalization of the nuclear density function (0.25\%)
and the precision estimated by the numerical integration, we conclude that our calculations
associated with the $^{12}$C form factors have a precision better that 1\% for $Q^2 \le 2$ GeV$^2$,
and that the calculations have a precision better than 0.5\% and 1\% for $^{16}$O and $^{40}$Ca,
respectively, for $Q^2 \le 2.5$ GeV$^2$.
The present precision can be improved by choosing grids with a few more points.
However, the present precision is sufficient for the present study and to quantify the difference
between the two calculations.

\begin{table}[t]
  \begin{tabular}{l |c  c c c  c c  c   c}
\hline
\hline
%& & &  Double ratios &\\
&   &  & DR  && & &\\
$Q^2$ (GeV$^2$)&  
0.5  & 1.0   &  1.5  & 2.0   & 2.5   \\
\hline
$\rho=0$  & \sp 1.000 \sp &  \sp 1.000 \sp  &  \sp 1.000  \sp & 1.000 \sp & \sp 1.000 \\
$^{12}$C  &  0.858 & 0.837 & 0.815 & 0.790 & 0.763\\
$^{16}$O  & 0.841 & 0.818 & 0.793 & 0.766 & 0.736 \\
$^{40}$Ca &  0.825 & 0.800 & 0.772 & 0.743 & 0.710\\
\hline
\hline
\end{tabular}
\caption{\footnotesize
  Model calculations for the double ratios (DR).}
\label{tab-radius}
\end{table}

\begin{table}[t]
  \begin{tabular}{l c|  c  c c c c}
\hline
\hline
  &  &  & &  $Q^2$ (GeV$^2$) & &  \\
   &   & \spQ  0.5 &  \spQ \spQ 1.0 &  1.5  &  2.0 \spQ \spQ & 2.5   \\
\hline
               &  $^{12}$C & \spQ 0.6 &  \spQ \spQ 1.5 & 2.4 & 3.3  \spQ \spQ & 4.3 \\
     $G_E^\ast$ &  $^{16}$O & \spQ 1.1  & \spQ \spQ 2.3 & 3.6 & 4.9 \spQ \spQ & 6.3\\
               & $^{40}$Ca & \spQ 1.4  & \spQ \spQ 3.0 & 4.6  & 6.3 \spQ \spQ  &  8.1\\
&   & \\
               &  $^{12}$C& \spQ 0.6  & \spQ \spQ 1.2 & 1.8 & 2.3 \spQ \spQ & 2.6\\
   $G_M^\ast$  &  $^{16}$O & \spQ 0.9  & \spQ \spQ 1.9 & 2.7 &  3.3 \spQ \spQ  & 3.8\\
              & $^{40}$Ca & \spQ  0.5 & \spQ \spQ 1.6 & 2.4 &  3.1\spQ \spQ  & 3.6\\
&   & \\
                  &  $^{12}$C  & \spQ 0.1 & \spQ \spQ 0.3 & 0.6 & 1.1 \spQ \spQ & 1.7\\
$G_E^\ast/G_M^\ast$ &  $^{16}$O  & \spQ 0.1 & \spQ \spQ 0.4 & 0.9 & 1.6 \spQ \spQ & 2.5\\
                   & $^{40}$Ca & \spQ 0.8 & \spQ \spQ 1.4 & 2.2 & 3.1 \spQ \spQ & 4.4\\
\hline
\hline
\end{tabular}
\caption{\footnotesize
Relative differences in percentage between the effective bound proton
form factors  $\overline{G_\ell^\ast} (Q^2)$ and the mean-field approximation  $G_\ell^\ast
\left(Q^2,\bar \rho_A \right)$.
\label{table-diffs}}
\end{table}

To improve the precision of the calculation, it is necessary to increase the precision of the
densities $\rho$, including more digits in the calculation of $\bar \rho_i$ particularly for small
densities (fractions of 0.01 $\rho_0$).

For a more systematic study of the EMFFs of the bound proton, 
we suggest, however, calculations based on smaller grids for $Q^2$, for the densities, and
interpolation methods if necessary. These estimates quantify the numerical integration uncertainty only. They do not include uncertainties associated with the density-dependent form factor parametrization, the nuclear density distributions, or the approximate correspondence with polarization-transfer observables.

\subsection{Double ratios for the bound neutron}

We use the opportunity to estimate also the EMFF double ratios of bound neutrons in
nuclei $^{12}$C,  $^{16}$O and $^{40}$Ca, since the formalism from
Refs.~\cite{Symmetry2025,Medium1} is extendable for all octet baryons.
The calculations for the effective form factor $\overline{G_\ell^\ast} (Q^2)$ and the mean-field
approximation are based on the expressions used for the proton, except for the form factors
$G_\ell^\ast \left(Q^2,\rho_i \right)$ that are now associated with the neutron, instead of the
proton.

The present estimates can be compared with data from the polarization-transfer method for the
neutron in the future in case the planned experiments become
possible~\cite{JLabPAC35}.
In the absence of experimental data, the model calculations can be used to infer the expected
behavior for the EMFFs of the bound neutron, and the differences for the case of the proton.

In the present case, we restrict the model calculations to the region $Q^2 \ge 0.5$ GeV$^2$.
Because the model is very sensitive to the parametrization of the meson cloud relevant in the
low-$Q^2$ region, we focus here more on the valence quark contributions, and
also because the framework is less well constrained for the case of the neutron.
The poor constraint of the parametrization for the neutron EMFFs is related to limitations in
the accuracy of the lattice QCD data for the neutron and neutral baryons~\cite{Medium1,Octet}.

\begin{figure}[t]
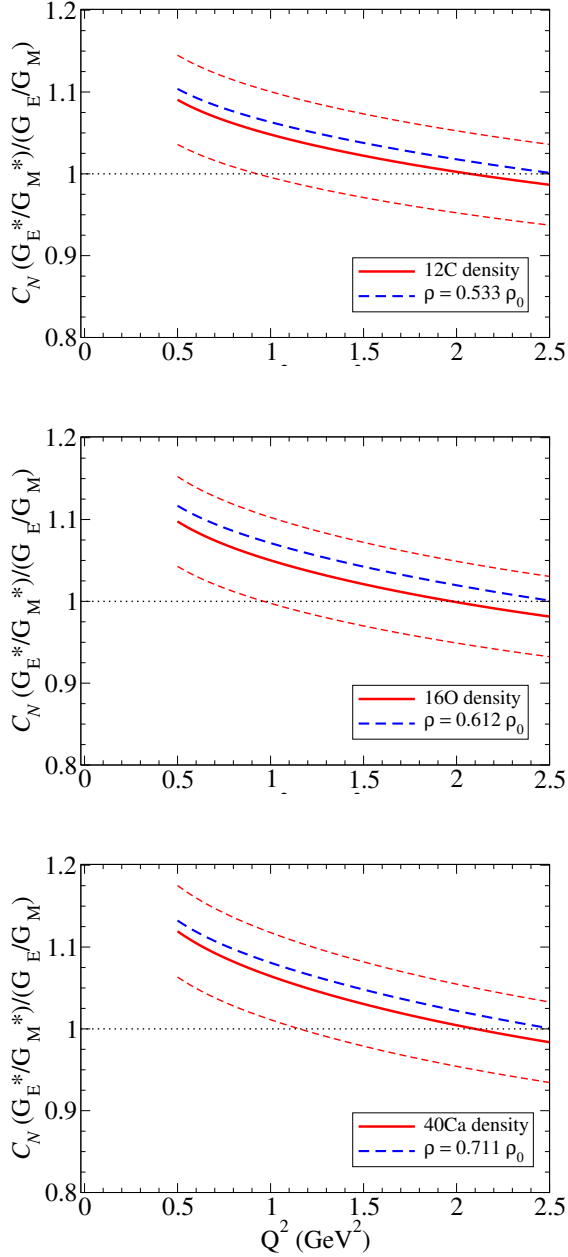
     %%  Figure 5 (page 10)
  %\vspace{.5cm}
  \begin{center}
    \includegraphics[width=2.9in]{NeutronDR-12C}
  \end{center}
  \begin{center}
    \includegraphics[width=2.9in]{NeutronDR-16O}
    \end{center}
  \begin{center}
    \includegraphics[width=2.9in]{NeutronDR-40Ca}
  \end{center}
  \caption{\footnotesize
Double ratios of the bound neutron in nuclei
$^{12}$C, $^{16}$O and $^{40}$Ca.
Labels as in Figs.~\ref{fig-12C}--\ref{fig-40Ca}.
As for the proton, the thin dashed curves indicate a variation of $\pm 5$\%
on the ratio calculations based on the density-profile function (solid line).
\label{fig-neutron} }
\end{figure}

The numerical calculations of the double ratios associated with the neutron EMFFs for the
nuclei $^{12}$C, $^{16}$O and $^{40}$Ca are presented in Fig.~\ref{fig-neutron}.
In the figures, we include the calculation associated with the effective form factors
$\overline{G_\ell^\ast} (Q^2)$ (solid line) and the mean-field approximation $G_\ell^\ast
\left(Q^2,\bar \rho_A \right)$ (dashed line).
As for the case of the proton, we include a 5\% band of variation for the effective form factors
to take into account the theoretical uncertainties on the determination
of experimental EMFF double ratios [see Eq.~(\ref{eqRatioA})].
The present calculations for the average densities of $^{12}$C, $^{16}$O and $^{40}$Ca, extend
previous calculations for the densities $\rho = 0.5 \rho_0$ and $\rho =\rho_0$~\cite{Medium1,Medium2}.
The results for $\rho = 0.5 \rho_0$ are very close to the mean-field estimates for
$^{12}$C ($\rho = 0.533 \rho_0$). 

From the observation of Fig.~\ref{fig-neutron}, we can conclude that the neutron double ratio is
expected to be enhanced in the nucleus in the low-$Q^2$ region, and reduced (quenching) above a
certain value of $Q^2$.
When we take into account the nuclear density functions (solid line)
we expect an enhancement up to $Q^2 \approx 2$ GeV$^2$ for all the nuclei under study.
The conclusion is then that, contrary to the proton (double ratio suppressed in-nucleus), for the
neutron we expect an enhancement of $G_E^\ast/G_M^\ast$ up to values of $Q^2 \approx 2$ GeV$^2$.

The opposite trends obtained for the proton and neutron constitute a characteristic prediction of the present model. The proton double ratio is suppressed throughout the calculated range, whereas the neutron ratio is enhanced up to $Q^2 \approx 2$ GeV$^2$. This qualitative proton–neutron contrast may provide a useful test of the flavor and meson cloud structure of the in-medium form factors. The neutron results should, however, be regarded as exploratory because $G_E^n$ is small and is more sensitive to the meson cloud parametrization and to the limited constraints on neutral baryon form factors.

From Fig.~\ref{fig-neutron}, we can also conclude that the shape of the nuclear density has a
stronger impact than in the case of the proton. The difference between the calculations with
$\overline{G_\ell^\ast} (Q^2)$ and the mean-field approximation is about 1.5--2.0\%
for $Q^2 \simeq 1$ GeV$^2$, while the differences for the proton are smaller.

An interesting property of the results from  Fig.~\ref{fig-neutron}, is that the double ratio
functions are very similar for all the three nuclei.
The slopes of the functions are similar, and the transition to the quenching regime happens
almost at the same point ($Q^2 = 2$ GeV$^2$).

From the comparison between the two calculations (solid and dashed lines), we can also conclude
that the relative difference between the two calculations does not vary much with $Q^2$, and it is
about 1.5--2.0\% depending on the nucleus.
From this observation, one concludes that we can use the mean-field approximation to estimate the
double ratios of the neutron in the first approximation (only one density considered), and use then
a correction factor (reduction of 1.5--2.0\%) to obtain a good estimate of the result associated
with the nuclear density functions.
This procedure avoids the necessity of the calculations of form factors for several values of the
nuclear density $\rho_i$.

The present results demonstrate that there is all the interest in performing experiments that
provide information about the impact of the nuclear matter (many nucleon) effects on the
neutron EMFFs.

\section{Outlook and conclusions  \label{sec-conclusions}}

In the present work we calculated the electric and magnetic form factors of
bound protons in middle mass nuclei.
We consider three different cases: $^{12}$C,  $^{16}$O and $^{40}$Ca associated with increasing
nuclear average densities, in the range $\bar \rho_A =$(0.5--0.72)$\rho_0$.
The method used in the present study can be extended to other nuclei, including
heavy nuclei. The central purpose of this comparison is to assess the validity of the average-density approximation previously used in applications of the covariant spectator–QMC model. The local-density concept itself has precedents in earlier finite-nucleus QMC studies, while the present calculation quantifies the density-profile correction for nucleus-averaged proton and neutron EMFF ratios within the current framework.

The present work constitutes a global analysis of the medium modifications based on the local
nuclear density profile function,  instead of the fully
differential analysis of the missing momentum of the knockout proton.
We consider a nucleus-averaged modification based on the spatial density distribution,
without explicitly resolving the missing momentum or the individual proton bound states.

The calculations are performed combining the covariant spectator quark model with the quark-meson
coupling model, based on the quark-meson coupling model calculated nuclear density profile functions.
Overall, we observe a suppression of the EMFF ratio
$G_E^\ast/G_M^\ast$ compared with the ratio in free space (quenching of ratio).

We also compare the calculations of form factors $G_E^\ast$ and $G_M^\ast$ with the estimates of
the mean-field approximation where the density of the nucleus is defined by the constant average
nuclear density, and the position-dependent shape of the nuclear density function is ignored.
We find that the average-density approximation reproduces the density-profile calculations very accurately below $Q^2 \simeq$ 0.25 GeV$^2$.
At larger $Q^2$, the density-profile form factors fall more slowly because the relative importance
of low-density surface components increases. The effect is larger for $G_E^\ast$
than for $G_M^\ast$ and generally becomes more pronounced for nuclei with larger average densities.
In general, the EMFFs  estimated with the nuclear density distribution
function have slower falloffs with $Q^2$ (softer quenching effect) than estimates based on the
mean-field approximation.
These results are explained by the dominance of the low nuclear densities (surface effect), when
compared with form factors associated with the average density, that are dominated by intermediate
densities.

Our calculations for the proton EMFF double ratios are compared with the double ratios
estimated based on the polarization-transfer method, under the assumption that the kinematic
corrections associated with target coefficients are small.
The present estimates are compatible with the available low-$Q^2$ data within the indicative correspondence band, although a quantitative comparison requires a dedicated reaction calculation.
For these results contributed the combination of the different binding energy
states of the proton ($s$, $p$, etc.).

We expect that the present model calculations may be tested in the near future by accumulated data
on polarization-transfer for the ratio $P_x^\prime/P_z^\prime$ at MAMI in the
low-$Q^2$ region and at Jefferson Lab for larger values of $Q^2$, hopefully up to $Q^2 = 2.5$
GeV$^2$.

We present also preliminary estimates for the neutron EMFF double ratios above $Q^2 = 0.5$
GeV$^2$, in order to infer the expected differences to the case of the proton.
In this case, we observe the opposite effect, the enhancement of the $G_E^\ast/G_M^\ast$ ratio up
$Q^2 \approx 2$ GeV$^2$.
We conclude also that in the case of the neutron double ratio, the mean-field approximation
overestimates the calculations that take into account the nuclear density by about 1.5–2.0\%.

\begin{acknowledgments}
G.R.~and M.-K.C.~were supported by the National
Research Foundation of Korea (Grant No.~RS-2021-NR060129).
K.T. was supported by the RCNP
Collaboration Research Network Program under Project No.~COREnet 057,
by the National Council for Scientific and Technological Development – CNPq, Brazil, Processes
No.~304199/2022-2 and No.~306866/2026-9, by the S\~{a}o Paulo Research Foundation (FAPESP),
Process No.~2023/07313-6 and  No.~2026/01656-7, and by the Instituto Nacional de Ci\^{e}ncia e
Tecnologia - Nuclear Physics and Applications (INCT-FNA), Brazil, Process No.~408419/2024-5.
\end{acknowledgments}

\appendix

\section{Numerical calculation of EMFF of the bound nucleons
  \label{appendix-p1}}

We discuss now the methods used in the numerical calculations of the EMFFs of
nucleon in a nucleus $A$.
Instead of considering the exact form for the nuclear density $\rho(r)$,
we consider an approximated function, hereafter mentioned as $\tilde \rho$.
The use of an approximated function is useful to make the connection with the model calculations
based on nuclear mediums with finite density, and it helps to obtain an intuitive estimate of the
dominant densities in the calculation of the EMFFs of the bound nucleons.

\begin{figure*}[t]
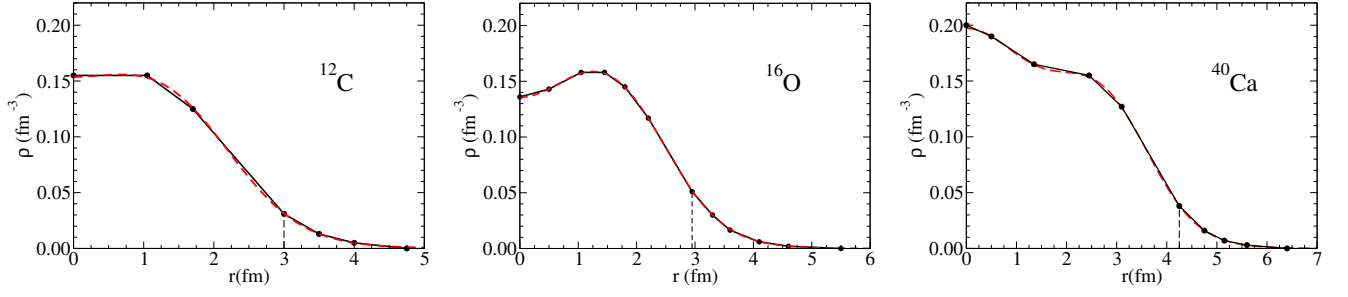
    %%      Figure 6
\vspace{.3cm}
\begin{center}
\mbox{
  \includegraphics[width=2.2in]{rho2-12C} \hspace{.1cm}
  \includegraphics[width=2.2in]{rho2-16O} \hspace{.1cm}
\includegraphics[width=2.2in]{rho2-40Ca} }
\end{center}
  \caption{\footnotesize
Approximated density functions $\tilde \rho(r)$ used the numerical calculations.
The vertical dashed line indicates the beginning of the region defined as the tail. $^{12}$C: 7
points; $^{16}$O: 12 points; $^{40}$Ca: 10 points.
\label{fig-rho-r2}}
\end{figure*}

\begin{table*}[t]
    \begin{tabular}{l l c c c c c c c c c c c c c c}
%\hline
%\hline
      &  &  $R_1$ & $R_2$ & $R_3$  & $R_4$ & $R_5$ & $R_6$ & $R_7$ \\
\hline
\hline
$^{12}$C & $R_i$  &   0.00 & 1.05 &  1.70 & 3.00 & 3.50  & 4.00 & 4.75\\
     & $\rho_i$ & 1.03 & 1.03 & 0.83 &  0.21 &  0.09 &  0.03 & 0.00\\ 
\hline
\hline
&   &  &  \\
      &  &  $R_1$ & $R_2$ & $R_3$  & $R_4$ & $R_5$ & $R_6$ & $R_7$ 
      & $R_8$   & $R_9$ & $R_{10}$& $R_{11}$  & $R_{12}$ \\
\hline
\hline
$^{16}$O & $R_i$  &  0.00 & 0.50 & 1.05 & 1.45 & 1.80 & 2.20 &  2.95 & 3.30 & 3.60 & 4.10 & 4.60 & 5.50\\
& $\rho_i$ & 0.91 & 0.95 & 1.05 & 1.05 &  0.97&  0.78 & 0.34 & 0.20 & 0.11
& 0.04 & 0.01 & 0.00\\ 
\hline
\hline
&   &  &  \\
&   &  $R_1$ & $R_2$ & $R_3$ & $R_4$ & $R_5$ & $R_6$ & $R_7$ 
      & $R_8$  & $R_9$ & $R_{10}$  \\ 
\hline
\hline
$^{40}$Ca & $R_i$  &   0.00 & 0.50 & 1.35 & 2.45 & 3.10 & 4.25 & 4.75 & 5.15 & 5.60 & 6.40 \\
     & $\rho_i$ & 1.33 & 1.33 & 1.10 & 1.03 & 0.85 & 0.25 & 0.11 & 0.05 & 0.02  &   0.00 &\\ 
\hline
\hline
  \end{tabular}
\caption{\footnotesize
  Nuclear density grids of the nuclei $^{12}$C,  $^{16}$O and $^{40}$Ca.
The radial variable ($R_i$) is in fm and densities and the densities $\rho_i$ are in units of $\rho_0$.
  \label{tab-grids}}
\end{table*}

\begin{table*}[t]
  %\vspace{5cm}
  \begin{center}
\begin{tabular}{l l}
\hline
\hline
Target  & \hspace{.7cm} Form factors\\
\hline
$^{12}$C &
$\overline{G_\ell^\ast} = 0.0627 \, G_\ell^\ast (1.03\rho_0) +
0.1806 \, G_\ell^\ast (0.93\rho_0) +  0.5366\, G_\ell^\ast (0.52\rho_0) 
 + G_{\ell, {\rm tail}}^\ast$   \\
 &  $ G_{\ell, {\rm tail}}^\ast = 0.1194 \, G_\ell^\ast (0.15\rho_0) +
 0.0651 \, G_\ell^\ast (0.06\rho_0) + 0.0356 \, G_\ell^\ast (0.02\rho_0) $ \\
 & \\
 %%%%%%%%%%%%%%%%%%%%%%%%%%%%%%%%%%%%%%%%%%%%%%%%%%%%%%%%%%%%%%%%%%%%%%%
 $^{16}$O &
$\overline{G_\ell^\ast} = 0.00463 \, G_\ell^\ast (0.93\rho_0) +
 0.0412 \, G_\ell^\ast (1.00\rho_0) +  0.0783\, G_\ell^\ast (1.05\rho_0) +
 0.1102\, G_\ell^\ast (1.01\rho_0)$ \\
 & $ \hspace{.7cm}  + \hspace{.1cm} 0.1642\, G_\ell^\ast (0.87\rho_0) +  0.3183\, G_\ell^\ast (0.56\rho_0) +  0.1079\, G_\ell^\ast (0.27\rho_0) 
 + G_{\ell, {\rm tail}}^\ast$   \\
 &  $ G_{\ell, {\rm tail}}^\ast = 0.0648 \, G_\ell^\ast (0.15\rho_0) +
 0.0643 \, G_\ell^\ast (0.07\rho_0) + 0.0292 \, G_\ell^\ast (0.03\rho_0)
 + 0.0170 \, G_\ell^\ast (0.01\rho_0)$ \\
 & \\
 %%%%%%%%%%%%%%%%%%%%%%%%%%%%%%%%%%%%%%%%%%%%%%%%%%%%%%%%%%%%%%%%%%
 $^{40}$Ca & $\overline{G_\ell^\ast} =
 0.00250 G_\ell^\ast (1.30\rho_0) + 0.0423  G_\ell^\ast (1.18\rho_0) +
 0.2027 G_\ell^\ast (1.07\rho_0)$ \\
  &   $ \hspace{.7cm}  + \hspace{.1cm}  0.2197 G_\ell^\ast (0.94\rho_0) + 
 0.3808 G_\ell^\ast (0.55\rho_0)  + G_{\ell, {\rm tail}}^\ast $ \\
 &  $ G_{\ell, {\rm tail}}^\ast =  0.0842  \, G_\ell^\ast (0.18\rho_0) +
 0.0348 \, G_\ell^\ast (0.08 \rho_0)
 + 0.0201 \, G_\ell^\ast (0.03\rho_0) +  0.0129 \, G_\ell^\ast (0.01\rho_0) $ \\
 \hline
\hline
\end{tabular}
\end{center}
\caption{\footnotesize
Analytic expressions used in the calculation of the effective in-medium form factors
$\overline{G_\ell^\ast} (Q^2)$, in terms of the constant density,
the form factors $G_\ell^\ast \left(Q^2,\bar \rho_i \right)$.
The variable $Q^2$ is omitted for simplicity.
\label{tab-FormFactors}}
\end{table*}

We consider then that the function $\rho(r)$ can be decomposed in a series of $n$ intervals:
$I_i = [R_i , R_{i+1} ]$ ($i = 1, .., n$),
where the function is approximated by a linear function
$\tilde \rho(r)$, and that that and that $\tilde \rho (r) =0$ for $r > R_{n+1}$.
The function  $\tilde \rho (r)$ is defined at the $n+1$ points $R_i$ by $\rho_i  = \tilde \rho(R_i)$.
By construction one has $R_1 = 0$, and $\rho_{n+1} = 0$.

In these conditions we can use the methodology from Sec.~\ref{secGell} and approximate
$\overline{G_\ell^\ast} (Q^2)$ by Eq.~(\ref{eqGl-medium2}),
where $\rho(r)$ is replaced by $\tilde
\rho (r)$, based on the (\ref{eqai}) and (\ref{eqZi}).
The grids used in the calculation of the integral $\overline{G_\ell^\ast} (Q^2)$ can be adapted to
the nucleus under discussion.
With this procedure, we can calculate the EMFFs of the
bound nucleon in $A$,
considering a sum of contributions associated with different densities, from the region near $r=0$
where $\bar \rho_i$ is close to the normal nuclear matter density $\rho_0$, to the region $\bar
\rho_i \simeq 0$ near the vacuum.

In the present study we consider the grids displayed in Fig.~\ref{fig-rho-r2}, where the functions
$\rho(r)$ are well approximated by linear functions in the intervals $[R_i , R_{i+1}]$.
The points associated to the figure are presented in Table~\ref{tab-grids}.

From the observation of Fig.~\ref{fig-rho-r2}, we can conclude that the shape of the nuclear
density profile function can be approximated by wider intervals from the region near $r=0$ up
to a value $\bar R$, where the function is reduced to about 20–25\% of the value of $\rho_1$.
For an accurate description of the region $r < \bar R$, we need, however, to consider intervals
with smaller length (about 0.5 fm).
The conclusion is that we can rewrite (\ref{eqGl-medium2}) in the form
\ba
\overline{G_\ell^\ast} (Q^2) =
\sum_{i=1}^m a_i G_\ell^\ast \left(Q^2, \bar \rho_i \right) +
G_{\ell \, {\rm tail} }^\ast (Q^2),
\label{eqGl-medium2R}
\ea
where the function $G_{\ell \, {\rm tail} }^\ast$ includes
$n-m$ intervals associated with the tail of the function $\rho (r)$.

The calculation of the term $G_{\ell \, {\rm tail} }^\ast$  requires some discussion.
Although the contribution of the tail can be small (small coefficients $a_i$), there is all the
convenience in including a sufficient number of points in order to ensure the accurate determination
of the integral, and the correct normalization of the function $\tilde \rho$ ($4 \pi {\cal Z} =A$).

The expressions used in the calculation of the form factors $\overline{G_\ell^\ast} (Q^2)$,
according to Eqs.~(\ref{eqGl-medium2}) and (\ref{eqGl-medium2R}), are presented in
Table~\ref{tab-FormFactors}.

At the end one needs to estimate the accuracy of the calculations based on the numerical
calculation of the integral $\overline{G_\ell^\ast} (Q^2)$.
The accuracy of the results can be estimated recalculating the expressions
with a larger number of points.

In the present work, we estimate the precision of our calculations, increasing the number of points
and using the algorithm associated with the Simpson method, as discussed in
Appendix~\ref{appendix-p2}.
In simple words, we extend the calculation based on the midpoint method (density $\bar \rho_i$)
with $n$ points associated with the average densities $\bar \rho_i$, with the inclusion of $n + 1$
endpoints of the intervals.
Recall that $n$ is the number of intervals chosen ($n + 1$ points). The precision of the
expressions used here ($n$ points/densities) is then tested with calculations that more $n + 1$
densities, with a total of $2n + 1$ densities.

\section{Numerical integration \label{appendix-p2}}

We consider two integration methods for the integration of linear functions $\tilde \rho (r)$: the
midpoint method, and the Simpson method.
The goal is the numerical calculation of the function $\tilde \rho (r)$ in the interval
$[R_i, R_{i+1} ]$:
\ba
{\cal I}_i = \int_{R_i}^{R_{i+1}} [r^2 \tilde \rho (r)] \,
G_\ell^\ast \left(Q^2, \tilde \rho (r) \right) \, dr,
\label{eq-int0}
\ea

In the following we use $\rho_i  = \tilde \rho (R_i)$, and
$\bar \rho_i = \sfrac{1}{2} (\rho_i + \rho_{i+1})$ to represent the average density.

When $G_\ell^\ast (Q^2, \rho)$ varies slowly over the density interval,
we approximate it by its value by the average density ${\bar \rho}_i$ (midpoint method),
and one can write
\ba
{\cal I}_i = {\cal Z}_i \, G_\ell^\ast (Q^2,\bar \rho_i),
   \label{eq-int1}
\ea
and
\ba
{\cal Z}_i \simeq \int_{R_i}^{R_{i+1}} r^2 \tilde \rho (r) \, dr.
\ea

When we consider the Simpson method, in addition to the midpoint density $\bar \rho_i$,
we also consider the endpoints $\rho_i$ and $\rho_{i+1}$.
Instead of (\ref{eq-int1}), we use
\ba
   {\cal I}_i =
   \left( \frac{1}{6}  G_\ell^\ast (Q^2, \rho_i) +
          \frac{4}{6}  G_\ell^\ast (Q^2, \bar \rho_i) +
          \frac{1}{6}  G_\ell^\ast (Q^2, \rho_{i+1}) 
   \right) {\cal Z}_i. \nonumber \\ 
\ea

In the midpoint method we calculate the form factors only for one density (midpoint density
$\bar \rho_i$).
In the Simpson method we use three densities: $\rho_i$, $\bar \rho_i$ and $\rho_{i+1}$.

\end{document}